\documentclass[aps,twocolumn,prd,10pt,showpacs,showkeys,preprintnumbers,superscriptaddress,floatfix,longbibliography,nofootinbib]{revtex4-1}

\pdfoutput=1
\usepackage{amsmath,amsfonts,amssymb,mathrsfs,graphicx,color,longtable}
\usepackage{hyperref}
\usepackage{graphicx}
\usepackage{amsfonts,amsmath,amssymb,bm}
\usepackage{array}
\usepackage{color}
\usepackage{enumitem}
\usepackage[explicit]{titlesec}
\usepackage[usenames,dvipsnames,table]{xcolor}
\usepackage[utf8]{inputenc}
\usepackage[letterpaper, portrait, margin=0.55in]{geometry}
\allowdisplaybreaks

\definecolor{nicered}{rgb}{0.7,0.1,0.1}
\definecolor{nicegreen}{rgb}{0.1,0.5,0.1}
\hypersetup{colorlinks,citecolor= nicegreen,linkcolor= nicered}

\newcommand{\beq}{\begin{equation}}
\newcommand{\eeq}{\end{equation}}
\newcommand{\bea}{\begin{eqnarray}}
\newcommand{\eea}{\end{eqnarray}}

\begin{document}

\def\Napoli{Dipartimento di Fisica, Universit{\`a} di Napoli Federico II and INFN, Sezione di Napoli, I-80126 Napoli, Italy}
\def\Zurich{Physik Institut, Universit{\"a}t Z{\"u}rich, CH-8057 Z{\"u}rich, Switzerland}
\def\MPI{Max-Planck-Institut f{\"u}r Physik, F{\"o}hringer Ring 6, 80805 M{\"u}nchen, Germany}
\def\Milano{Universit{\`a} di Milano-Bicocca and INFN, Sezione di Milano-Bicocca, Piazza della Scienza 3, 20126 Milano, Italy}

\preprint{MPP-2020-69, ZU-TH 14/20}

\title{Lepton-quark collisions at the Large Hadron Collider}

\author{Luca Buonocore}
\email[Electronic address:]{luca.buonocore@na.infn.it} 
\affiliation{\Napoli}
\affiliation{\Zurich}
\author{Ulrich Haisch}
\email[Electronic address:]{haisch@mpp.mpg.de} 
\affiliation{\MPI}
\author{Paolo Nason}
\email[Electronic address:]{paolo.nason@mib.infn.it} 
\affiliation{\Milano}
\author{Francesco Tramontano}
\email[Electronic address:]{francesco.tramontano@cern.ch}
\affiliation{\Napoli}
\author{Giulia Zanderighi}
\email[Electronic address:]{zanderi@mpp.mpg.de} 
\affiliation{\MPI}

\begin{abstract}
Processes commonly studied at the Large~Hadron~Collider~(LHC) are induced by quarks and gluons inside the protons of the LHC beams. In this letter we demonstrate that, since protons also contain  leptons, it is possible to target lepton-induced processes at the LHC as well. In particular, by picking a lepton from one beam and a quark from the other beam, we present for the first time a comprehensive analysis of resonant single leptoquark~(LQ) production at a hadron collider.  In the case of minimal  scalar LQs, we derive novel bounds  that arise from the LHC~Run~II considering all possible flavour combinations of an electron or a muon and an up ($u)$, a down ($d$), a strange or a~charm quark.  For the flavour combinations with a $u$ or a $d$ quark, the obtained limits represent the most stringent constraints to date  on LQs of this type. The prospects of our method at future LHC runs are also explored. Given the discovery reach of the proposed LQ signature, we argue that dedicated resonance searches in final states featuring a single light lepton and a single light-flavour jet should be added to the exotics search canon of both the ATLAS and the CMS collaboration.
\end{abstract}


\maketitle

{\bf Introduction.} At the Large Hadron Collider (LHC) an immense number of collisions between quarks and gluons took place and many more are expected in the upcoming Run~III and the high-luminosity (HL-LHC) upgrade. These collisions have been studied in great detail and have been used to perform precision measurements of Standard~Model~(SM) processes and to search for physics beyond the~SM~(BSM). Due to quantum fluctuations, protons however also contain charged leptons, making it possible to study lepton-induced processes at the LHC as well. The simplest process of this kind consists  of the collision between a lepton ($\ell$) from one proton  and a quark ($q$) from the other proton, giving rise to the resonant production of an exotic leptoquark~(LQ) state. LQs are hypothetical coloured bosons that carry both a baryon number and a lepton number~\cite{Pati:1974yy} --- see~\cite{Dorsner:2016wpm} for a recent LQ review. Below, taking the case of LQs as an example, we  demonstrate that lepton-induced  processes, which so far have been completely  neglected at the LHC, can be complementary to quark- or gluon-induced channels in searches for BSM physics. 

A number of different searches for LQs have been considered so far at the LHC. For example, LQs can be pair-produced at hadron colliders via  quark-fusion $\big($i.e.~$q \bar q \to {\rm LQ} \, \overline{\rm LQ} \to (\ell q) (\ell q)$$\big)$ or gluon-fusion $\big($i.e.~$g g \to {\rm LQ} \, \overline{\rm LQ} \to (\ell q) (\ell q) $$\big)$. For sufficiently large LQ-lepton-quark couplings, also pair-production via $t$-channel exchange of a lepton becomes relevant, and LQ exchange contributes to Drell-Yan like dilepton production $q \bar q \to \ell^+ \ell^-$ and single LQ production in $g q \to \ell^+ \ell^- q$. Both the ATLAS and the CMS collaborations have exploited the different channels  to constrain the parameter space of LQ models (cf.~\cite{Aaboud:2016qeg,Aaboud:2019jcc,Aaboud:2019bye,Khachatryan:2016jqo,Sirunyan:2018ryt,Sirunyan:2018vhk,Sirunyan:2018btu} for the latest results) and the subject has also received renewed theoretical interest~(see~for~instance~\cite{Mandal:2015vfa,Faroughy:2016osc,Raj:2016aky,Diaz:2017lit,Dorsner:2018ynv,Hiller:2018wbv,Bansal:2018eha,Angelescu:2018tyl,Schmaltz:2018nls,Mandal:2018kau,Baker:2019sli,Chandak:2019iwj,Padhan:2019dcp,Bhaskar:2020gkk}).

\begin{figure}[!t]
\begin{center}
\vspace{2mm} 
\includegraphics[width=0.4\textwidth]{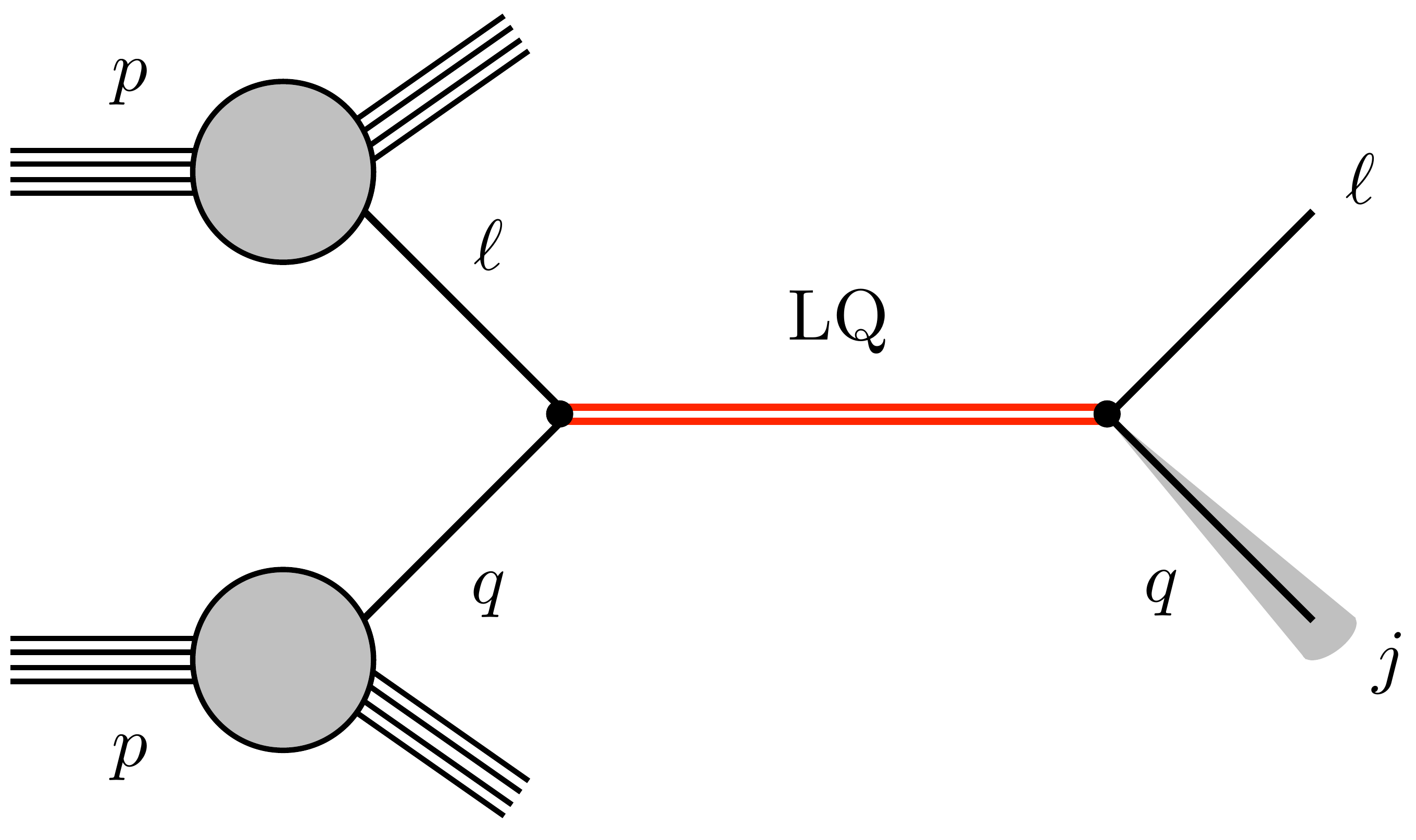} 
\vspace{4mm} 
\caption{\label{fig:diagram} Tree-level Feynman diagram describing resonant single LQ production in $pp$ collisions with a final state consisting of  a lepton  and a jet. }
\end{center}
\end{figure}

In this letter,  we  develop a new LQ search strategy, which relies on the fact that once QED interactions are considered, lepton parton distribution functions~(PDFs)  of the proton~($p$) are generated.  The lepton PDFs are small compared to  those of quarks and gluons, since they are suppressed by two powers of the ratio of the electromagnetic coupling constant over the strong coupling constant. A~first implementation of a PDF set with leptons was put forward in~\cite{Bertone:2015lqa}, that was however affected by large uncertainties.  Based upon the so-called {\tt LUX} method~\cite{Manohar:2016nzj,Manohar:2017eqh}, a precise determination of the lepton PDFs  has become available recently~\cite{Buonocore:2020nai}. In the following we will use the latter lepton PDF determination to compute the cross sections for resonant single LQ production $\ell q \to {\rm LQ}  \to \ell q$  in $pp$ collisions, studying final states with a  high transverse momentum~($p_T$) electron~($e$) or muon~($\mu$) and a high-$p_T$  light-flavour  jet~($j$),  to derive bounds on the parameter space of   minimal scalar LQs  for all possible flavour combinations involving first-  and second-generation leptons and quarks.\footnote{The same idea has already been brought forward more than 20~years ago in \cite{Ohnemus:1994xf,Eboli:1997fb}, but a reliable estimate of the lepton PDFs was missing, and the resulting LHC phenomenology has never been worked out in detail.}  The corresponding tree-level Feynman diagram is shown in~Fig.~\ref{fig:diagram}. 

{\bf LQ interactions.}  We  follow~\cite{Schmaltz:2018nls} and consider scalar LQs which couple  only to one lepton and quark flavour, taking them to be singlets under the $SU(2)_L$ part of the SM gauge group. In order to obtain $SU(2)_L$ invariant interactions, we couple the LQs to the $SU(2)_L$ singlet leptons and quarks, i.e.~the right-handed SM fermions. Using the notation where all singlet fields are represented by left-handed charge conjugate fields, the scalar LQ  coupling to singlet electrons and up quarks can then be written as 
\beq \label{eq:lagrangian}
{\cal L} \supset \lambda_{eu} \hspace{0.25mm} {\rm LQ}_{eu} \left ( E^c U^c \right)^\ast + {\rm h.c.} \,,
\eeq
where the spinor indices of $E^c$ and $U^c$ are contracted anti-symmetrically. In the limit of large scalar LQ masses,~i.e.~$M  \gg m_\ell, m_q$, the corresponding 
total decay width of the LQ is given by 
\beq \label{eq:decaywidth}
\Gamma  = \frac{|\lambda_{eu}|^2}{16 \pi} \hspace{0.5mm} M \,,
\eeq
 and due to the minimal character of the LQ, it decays almost exclusively  to  final states with an electron and a up quark. Notice that decays to anti-particles such as ${\rm LQ} \to e^+ u$ are not induced by~(\ref{eq:lagrangian}). The expressions~(\ref{eq:lagrangian}) and~(\ref{eq:decaywidth}) also apply to all other flavour combinations after obvious replacements of fields and indices. 

{\bf Analysis strategy.} The signal predictions corresponding to $s$-channel single LQ production $pp \to {\rm LQ} \to \ell q$ are generated at leading order~(LO) using the implementation of the Lagrangian~(\ref{eq:lagrangian}) presented in~\cite{Dorsner:2018ynv} together with  the {\tt LUXlep}~PDF set, which has been obtained by combining the lepton PDFs of~\cite{Buonocore:2020nai} with the  {\tt NNPDF3.1luxQED} set~\cite{Bertone:2017bme}. Event generation and showering is  performed  with {\tt MadGraph5\_aMC\@NLO}~\cite{Alwall:2014hca} and {\tt PYTHIA~8.2}~\cite{Sjostrand:2014zea}. Since at present {\tt PYTHIA~8.2} cannot handle incoming leptonic partons, we have replaced all initial state leptons by photons in the Les~Houches files to shower the events.\footnote{This replacement leads to a   mismodelling of the signal strength after imposing the lepton and jet veto present in our analysis. We estimate this effect to be of ${\cal O}(10\%)$, and therefore to only mildly affect the derived LQ limits.}  The parton shower backward evolution of {\tt PYTHIA~8.2}  then produces only quarks from photon splitting, so after showering our simulation includes initial-state quarks instead of leptons.\footnote{We remark that the inclusion of incoming leptons in shower Monte~Carlos is not difficult to realise and one can expect it to become available in the near future~\cite{PRTS}.} 

Our analysis uses experimentally identified jets, electrons, muons and missing transverse energy~($E_{T, {\rm miss}}$).  {\tt FastJet~3}~\cite{Cacciari:2011ma} is used to construct anti-$k_t$ jets~\cite{Cacciari:2008gp} of radius $R = 0.4$. Our analysis is implemented in {\tt CheckMATE~2}~\cite{Dercks:2016npn}, which employs {\tt Delphes~3}~\cite{deFavereau:2013fsa} as a fast detector simulator.  Detector effects are  simulated by smearing  the momenta of the reconstructed objects, and by applying reconstruction and identification efficiency factors tuned to mimic the performance of the ATLAS detector. In particular, electron candidates are required to satisfy the tight identification criteria of ATLAS~\cite{Aad:2019tso}, while muon candidates must fulfil the ATLAS quality selection criteria optimised for high-$p_T$ performance~\cite{Aad:2016jkr,1793218}. The corresponding reconstruction and identification efficiency for electrons amounts to $90\%$ for  $p_T > 500 \, {\rm GeV}$, while for muons the reconstruction and identification efficiency is $69\%$ ($57\%$) at $p_T = 1 \, {\rm TeV}$ ($p_T = 2.5 \, {\rm TeV}$) --- cf.~for~example~\cite{Aad:2019fac,Aad:2019wvl}.  $E_{T, {\rm miss}}$  is reconstructed from the sum of the smeared calorimeter deposits, including an extra smearing factor that effectively parametrises additional QCD activity due to pile-up and has been tuned to match the ATLAS distributions.

The basic selections in our signal region require a  lepton~($e$ or $\mu$) with $|\eta_\ell| < 2.5$ and $p_{T, \ell} > 500 \, {\rm GeV}$ and a  light-flavour jet with  $|\eta_j| < 2.5$ and $p_{T, j} > 500 \, {\rm GeV}$.  We~furthermore demand $E_{T, {\rm miss}} < 50 \, {\rm GeV}$, veto events  that contain additional leptons with $|\eta_\ell| < 2.5$ and $p_{T,\ell} > 7 \, {\rm GeV}$ and impose a jet veto on subleading jets with $|\eta_j| < 2.5$ and $p_{T, j} > 30 \, {\rm GeV}$. The jet veto limits the amount of hadronic activity and ensures that  the background from $t \bar t$,  and $s$- and $t$-channel single top production are negligible in the signal region. 

The dominant background  turns out to be $W^- + j$ production which is generated at next-to-leading order~(NLO)~in QCD.\footnote{Assuming the same LQ production cross section and branching ratios the  limits following from $\ell^+ + j$ searches turn out to be practically  identical to those resulting from $\ell^- + j$ searches.   This feature can be understood by noting that after imposing the discussed selections the size and the shape of the $W^- + j$  and  $W^+ + j$ backgrounds are very similar. This is due to the fact that the charged lepton in the $W^+ \to \ell^+ \nu_\ell$ decay is softer than the one in the $W^-  \to \ell^- \bar \nu_\ell$ decay, which compensates the larger value of the $W^+ + j$ cross section originating from the larger up-quark content of the proton.} Next-to-next-leading order~QCD and electroweak effects that would effectively reduce the size of the  $W^- + j$ background prediction in the phase space region of interest~\cite{Lindert:2017olm} are not included in our analysis. Subleading backgrounds arise from $Z+j$, $WW$, $W^-Z$ and $tW$ production and are simulated at LO and normalised to the known NLO~QCD cross sections. At high values of $m_{\ell j}$ also  $\ell^- + j$ production from an initial-state lepton and quark via $t$-channel exchange of a photon or $Z$ boson represents a relevant irreducible background. We include this background at~LO.\footnote{As a check of the simulation we have computed the signal and backgrounds also using {\tt POWHEG}~\cite{Nason:2004rx,Frixione:2007vw,Alioli:2009je,Alioli:2010xd,Re:2010bp,Alioli:2010qp,Melia:2011tj}, and modelled detector effects by simple smearing of the particle momenta at the Monte~Carlo truth level, along the lines of~\cite{Buonocore:2020nai}.} For each background process, the number of events after cuts is fitted and extrapolated to high invariant lepton-jet masses $m_{\ell j}$  using $e^{-a} \hspace{0.5mm} m_{\ell j}^{b + c \ln  m_{\ell j}}$. This functional form is  commonly used in experimental searches (see~for~instance~\cite{Aad:2019wvl}).

\begin{figure}[!t]
\begin{center}
\vspace{2mm} 
\includegraphics[width=0.45\textwidth]{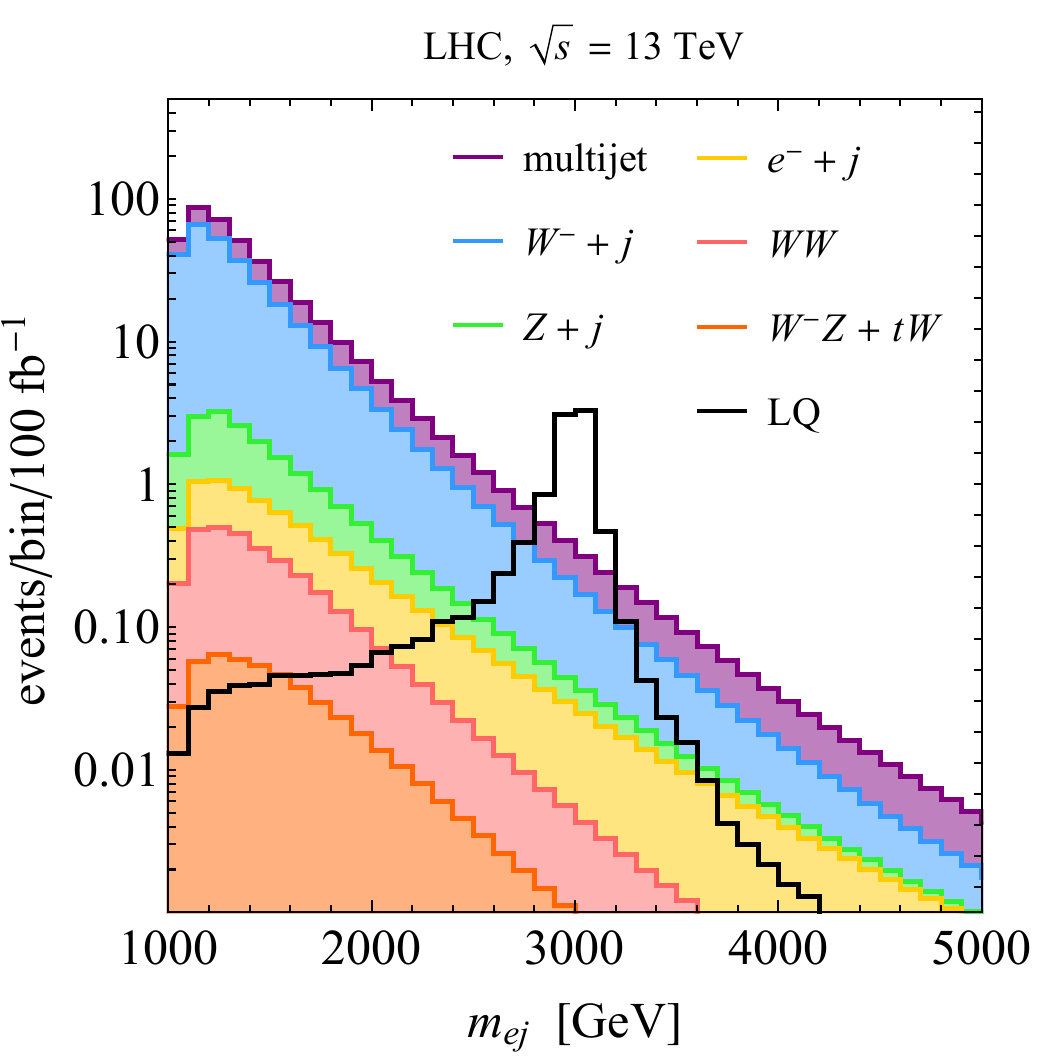} 
\vspace{0mm} 
\caption{\label{fig:background} Distributions of $m_{ej}$ after imposing the experimental selection requirements as detailed in the text. The coloured histograms are stacked and represent the SM backgrounds. The LQ signal prediction corresponds to $M  = 3 \, {\rm TeV}$ and $\lambda_{eu} = 1$ and is superimposed as a  black line. The events are binned in $100 \, {\rm GeV}$ bins and all predictions are obtained for $100 \, {\rm fb}^{-1}$ of $pp$ collisions at a centre-of-mass energy of $\sqrt{s} = 13 \, {\rm TeV}$.}
\end{center}
\end{figure}

In~actual experimental analyses the background contributions from events with fake or non-prompt electrons from the decay of heavy-flavour hadrons is typically extracted from the data using a matrix method as described for example in~\cite{Aaboud:2016zkn}. Since we cannot perform such a  data-driven background estimate, we mimic the impact of the multijet background in our analysis by taking  it from the ATLAS $\ell + j$ analysis~\cite{Aad:2013gma}. Using the auxiliary material of~\cite{Aad:2013gma} as provided in~\cite{BH}  we estimate that the ratio of the multijet to the other $e^- + j$ backgrounds amounts to $28\%$, $84\%$ and $142\%$ for $m_{ej}$ values of  $1 \, {\rm TeV}$, $3 \, {\rm TeV}$ and $5 \, {\rm TeV}$, respectively. 

In Fig.~\ref{fig:background} the distributions of the invariant mass~$m_{ej}$ of the electron and the leading jet are displayed for the SM~backgrounds, and  for a benchmark minimal scalar LQ with mass  $M = 3 \, {\rm TeV}$ and $\lambda_{eu} = 1$, after applying the event selection described above.  An integrated luminosity of $100  \, {\rm fb}^{-1}$ under LHC~Run~II conditions is assumed. Our benchmark LQ has a width of $\Gamma \simeq 60 \, {\rm GeV}$. One observes that the sum of the SM backgrounds is a steeply falling distribution, while the LQ signal exhibits a narrow  peak as indicated by the black line. The $t \bar{t}$,  and  $s$- and $t$-channel single top backgrounds are not shown, since they are very small.

Systematic uncertainties are treated as follows in our analysis. The~scale~(PDF) uncertainties affecting the dominant $W^- + j$ background have been determined with {\tt MadGraph5\_aMCNLO}. They amount to $4.6\%$ ($1.2\%$), $12\%$ ($4.2\%$) and $41\%$ ($57\%$) for~$m_{\ell j}$ values of $1 \, {\rm TeV}$, $3 \, {\rm TeV}$ and $5 \, {\rm TeV}$, respectively. No~systematic uncertainty is applied to the signal predictions.\footnote{We have checked that the scale (PDF) uncertainties of the signal amount to less than $10\%$ ($3\%$) for LQ masses in the range of $1 \, {\rm TeV}$  to $5 \, {\rm TeV}$. These uncertainties would hence affect the limits on the couplings $\lambda_{\ell q}$ at the few percent level only.}  The individual sources of uncertainty are added in quadrature which results in total systematic background uncertainties of  $4.7\%$, $13\%$ and $70\%$ for~$m_{\ell j}$ values of  $1  \, {\rm TeV}$, $3 \, {\rm TeV}$ and $5 \, {\rm TeV}$, respectively.

{\bf LHC constraints.}  The resonance line shape of the LQ signal  is modelled by a relativistic  Breit-Wigner  that is fitted to the distribution of  events after showering, reconstruction and cuts. In this way the broadening of the peak by PDF, parton shower and  non-perturbative effects is described.  We~find that  compared to (\ref{eq:decaywidth}) the peak  is broadened by the latter effects  by a factor of 6.0, 3.1 and 2.2  for a LQ~mass of $1 \, {\rm TeV}$, $ 3 \, {\rm TeV}$ and  $5 \, {\rm TeV}$, respectively. 

The statistical significance of any localised excess in the $m_{\ell j}$ distribution is quantified using a sliding window approach after binning the background and signal predictions. The bin size is thereby taken to be equal to the $m_{\ell j}$ resolution.This resolution is estimated by combining the information on the dilepton and dijet mass resolutions given in~\cite{Aad:2019fac,dilepton} and~\cite{Aad:2019hjw,dijet}, respectively. Using a simple error propagation, we find that in the electron case the experimental mass resolution amounts to $2.3\%$, $1.7\%$ and $1.6\%$ at $1 \, {\rm TeV}$, $3 \, {\rm TeV}$ and $5 \, {\rm TeV}$, respectively. In the muon case the corresponding numbers are $6.7\%$, $12\%$ and $17\%$. The width of the search window is then varied from a minimum of twice the $m_{\ell j}$ resolution up to $2 \, {\rm TeV}$, and the optimal width is determined for each signal hypothesis  such that the LQ signal  deviates most significantly from the smooth background distribution. The significance is calculated as a Poisson ratio of likelihoods modified to incorporate systematic uncertainties on the background using the Asimov approximation~\cite{cowan}.

\begin{figure*}[t]
\begin{center}
\vspace{2mm} 
\includegraphics[width=0.45\textwidth]{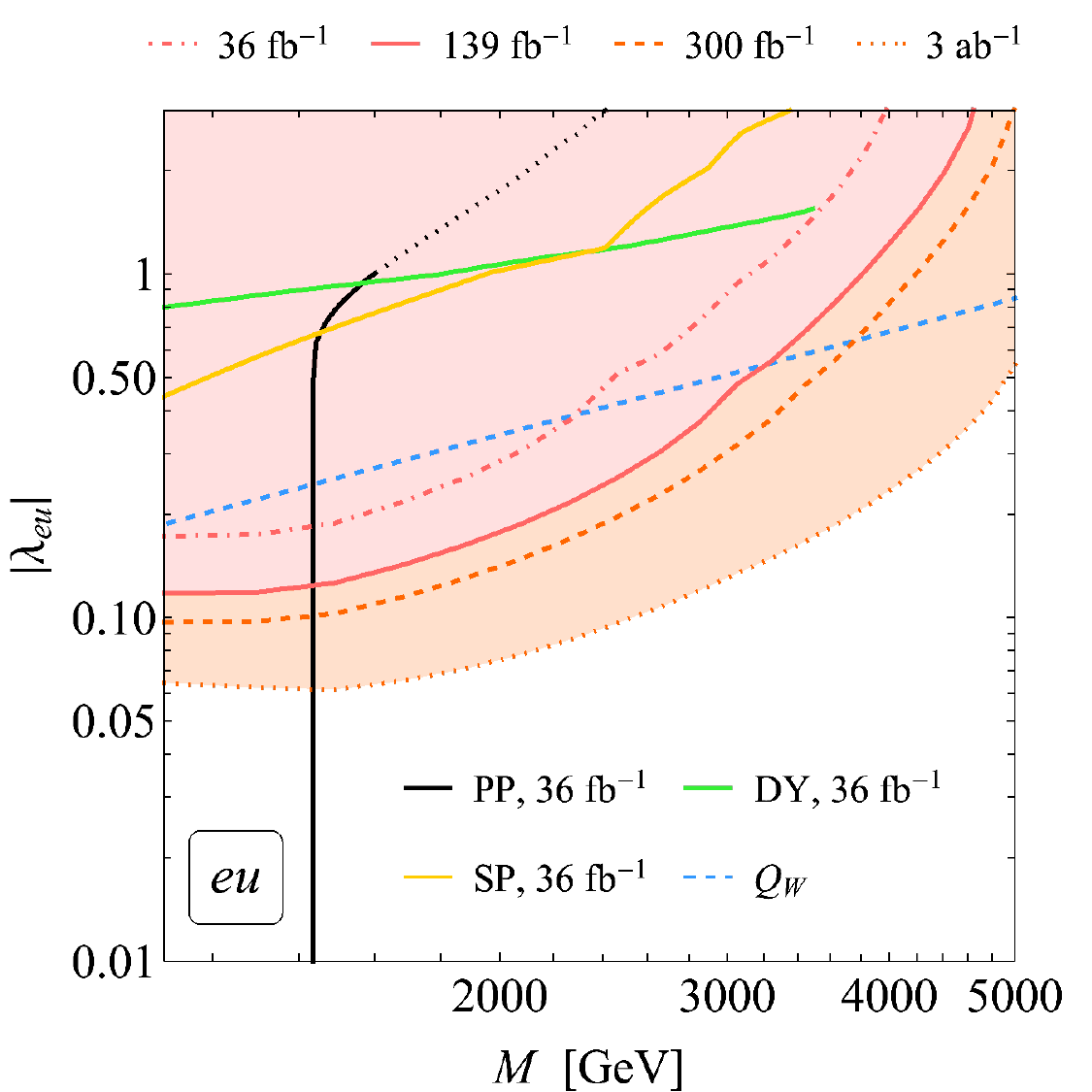} \qquad \includegraphics[width=0.45\textwidth]{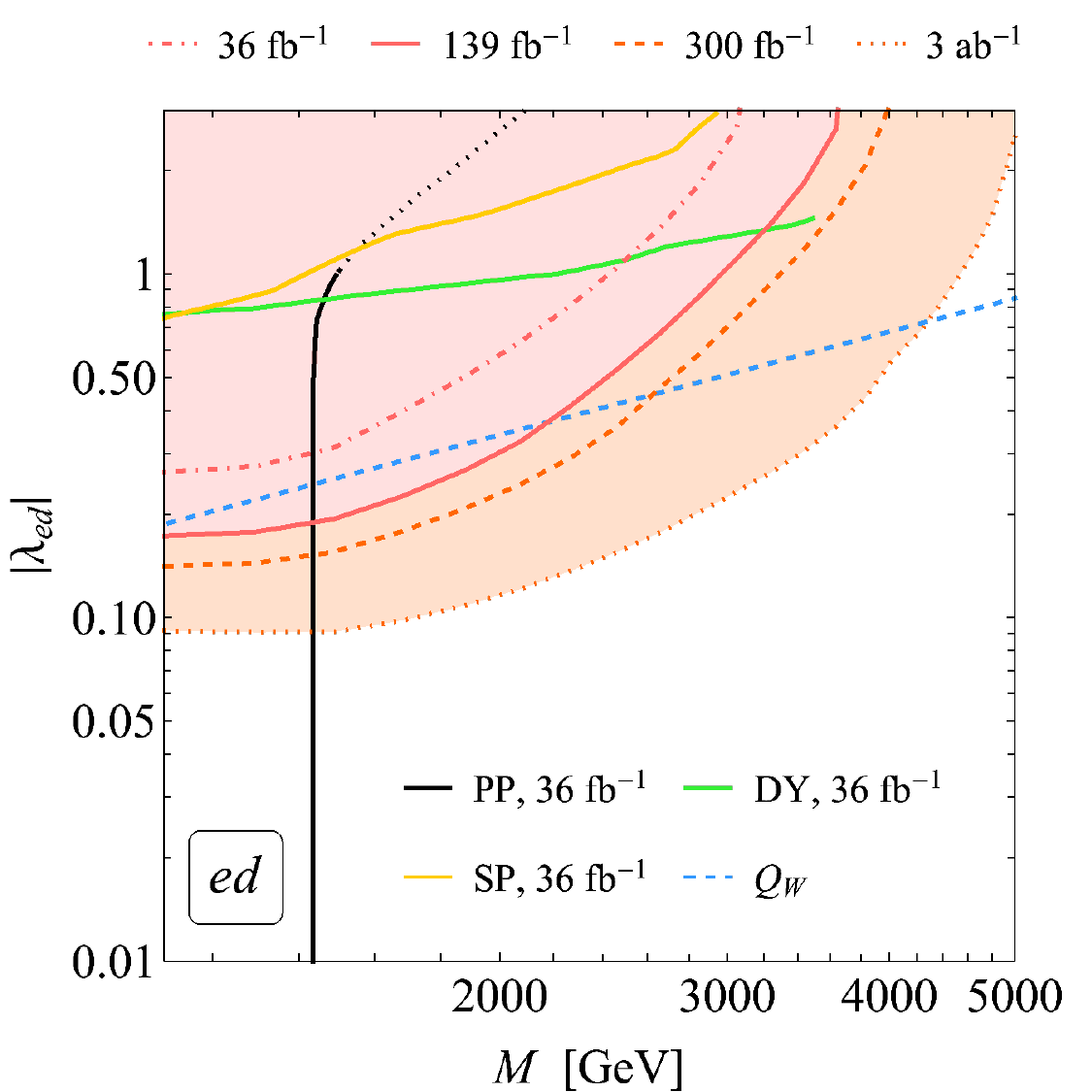} 

\vspace{4mm} 

\includegraphics[width=0.45\textwidth]{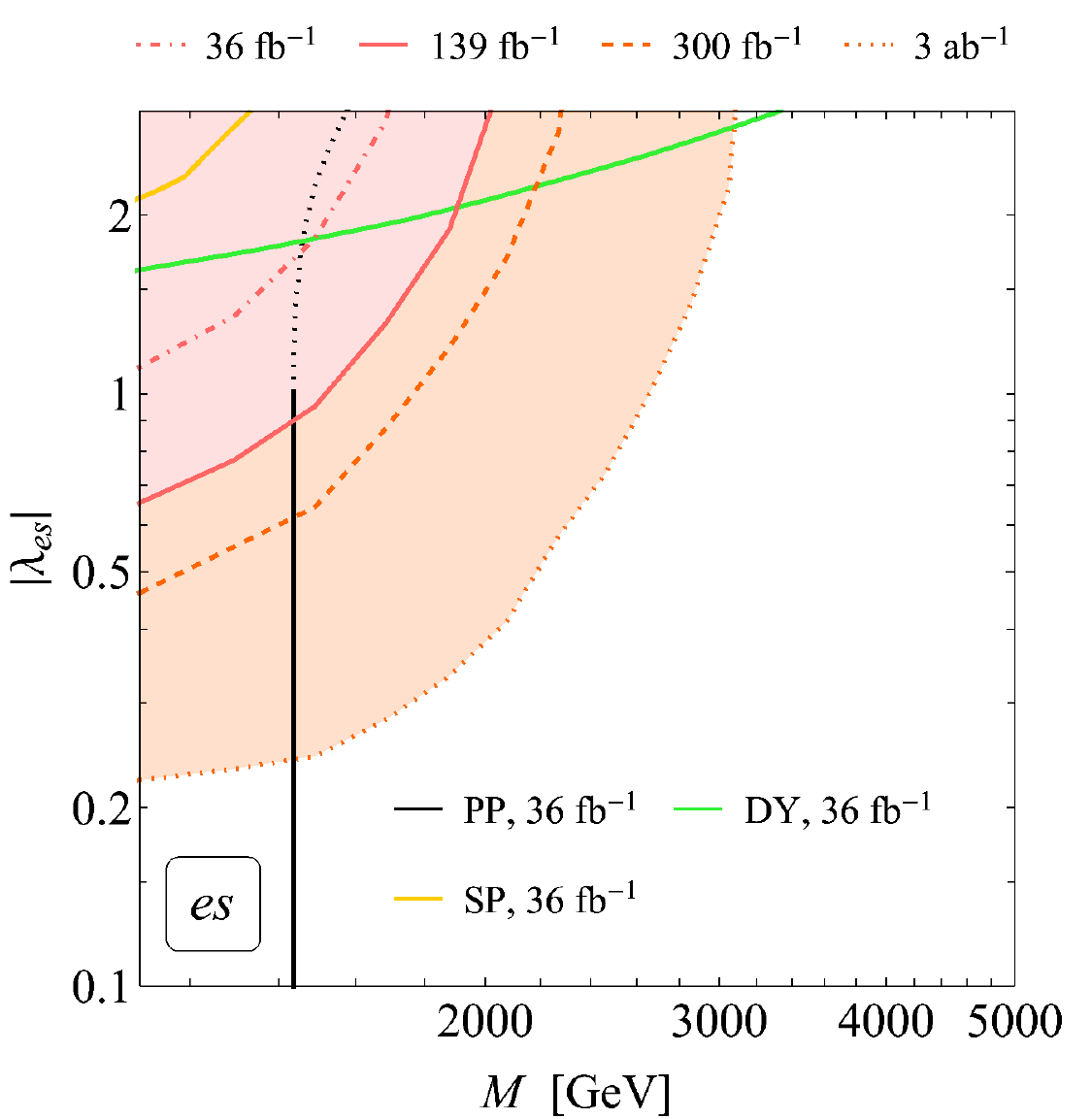} \qquad \includegraphics[width=0.45\textwidth]{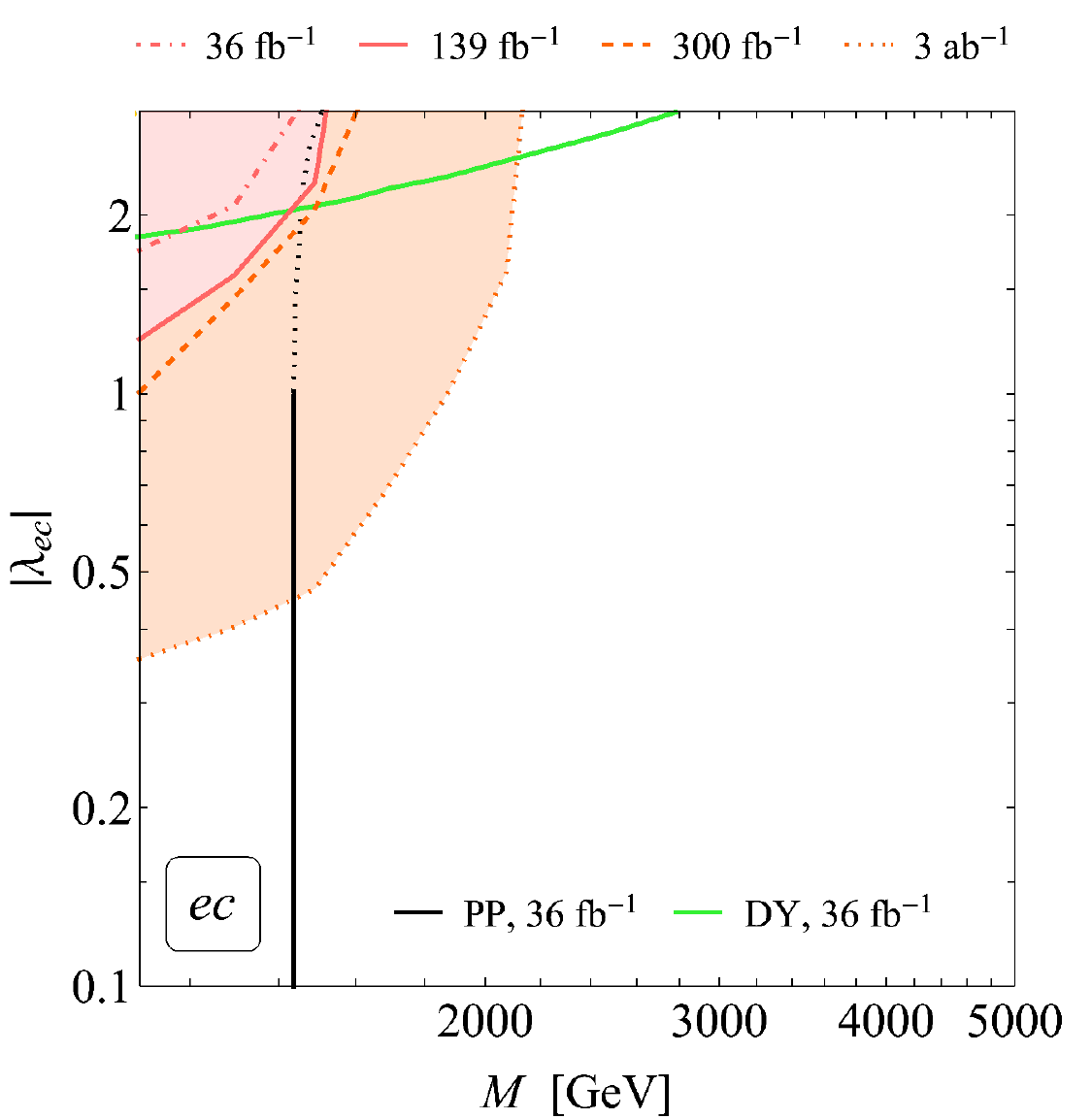} 
\vspace{2mm} 
\caption{\label{fig:limits1} 95\%~CL limits on the parameter space of minimal ${\rm LQ}_{e q}$ bosons with $q = u, d, s, c$. The red (orange) shaded regions correspond to the parameter space that is excluded by resonant single LQ production at the LHC~Run~II (future LHC runs). The black lines  indicate the PP limits obtained in~\cite{Schmaltz:2018nls} by recasting the results~\cite{Sirunyan:2018btu}, the green lines correspond to the  DY bounds derived in~\cite{Schmaltz:2018nls} using~\cite{Sirunyan:2018exx}, while the  yellow lines represent the SP projections~\cite{Schmaltz:2018nls} of  the search~\cite{Khachatryan:2015qda}. The dashed blue lines depicts the constraints from $Q_W$ measurements~\cite{Schmaltz:2018nls}.  The parameter spaces to the left and/or above the lines are ruled out. See text for further~details. }
\end{center}
\end{figure*}

\begin{figure*}[t]
\begin{center}
\vspace{2mm} 
\includegraphics[width=0.45\textwidth]{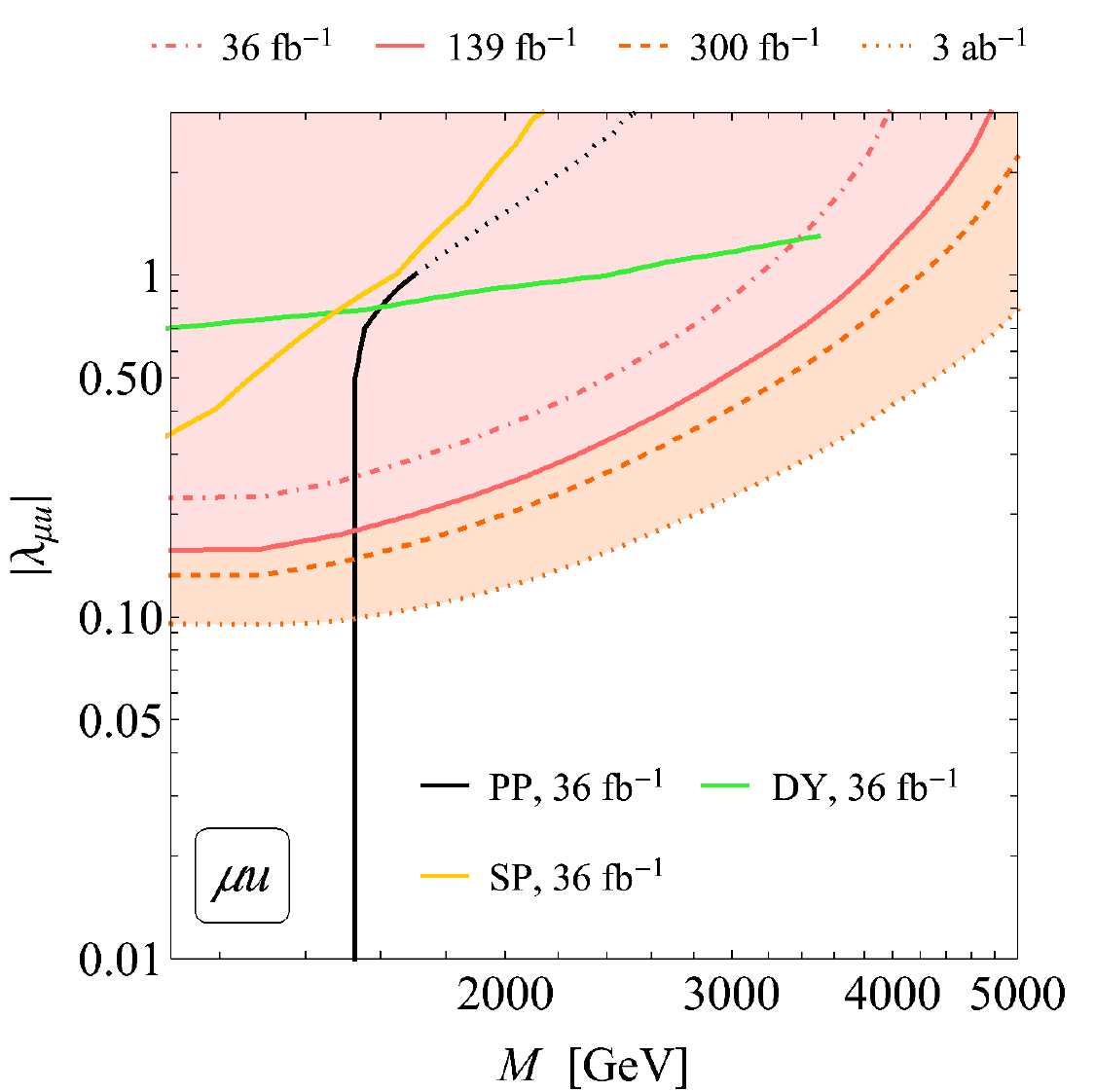} \qquad \includegraphics[width=0.45\textwidth]{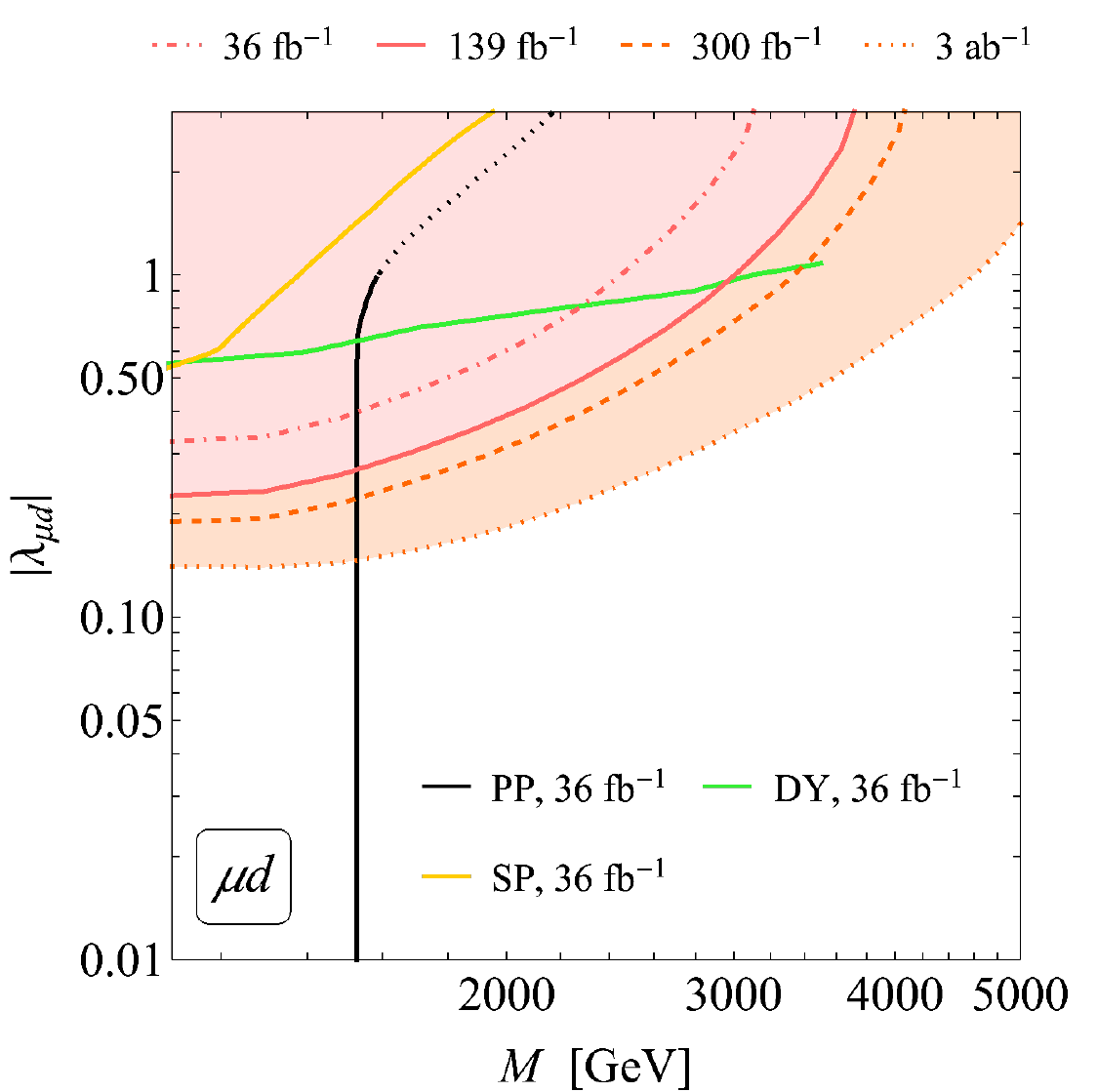} 

\vspace{4mm} 

\includegraphics[width=0.45\textwidth]{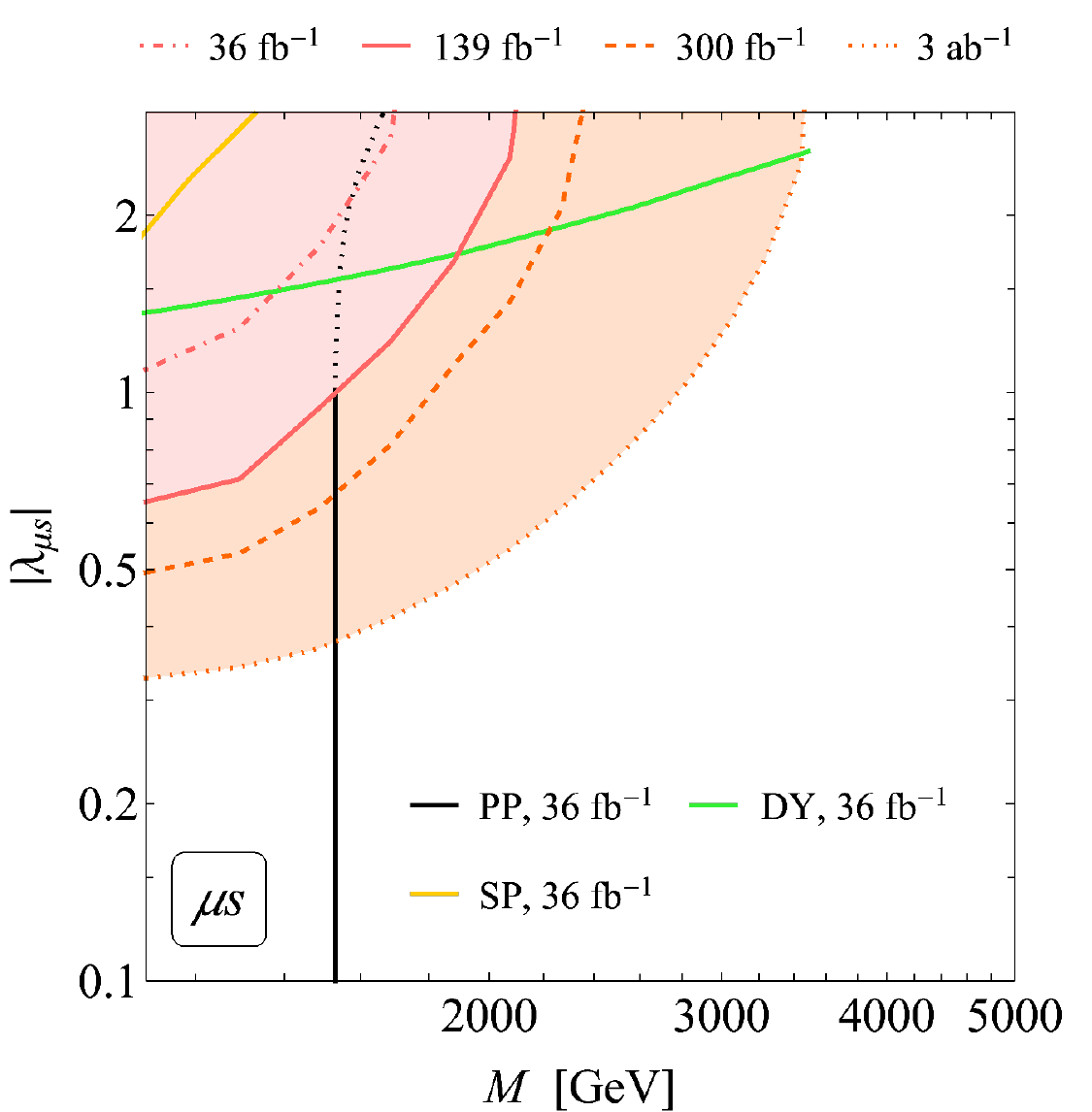} \qquad \includegraphics[width=0.45\textwidth]{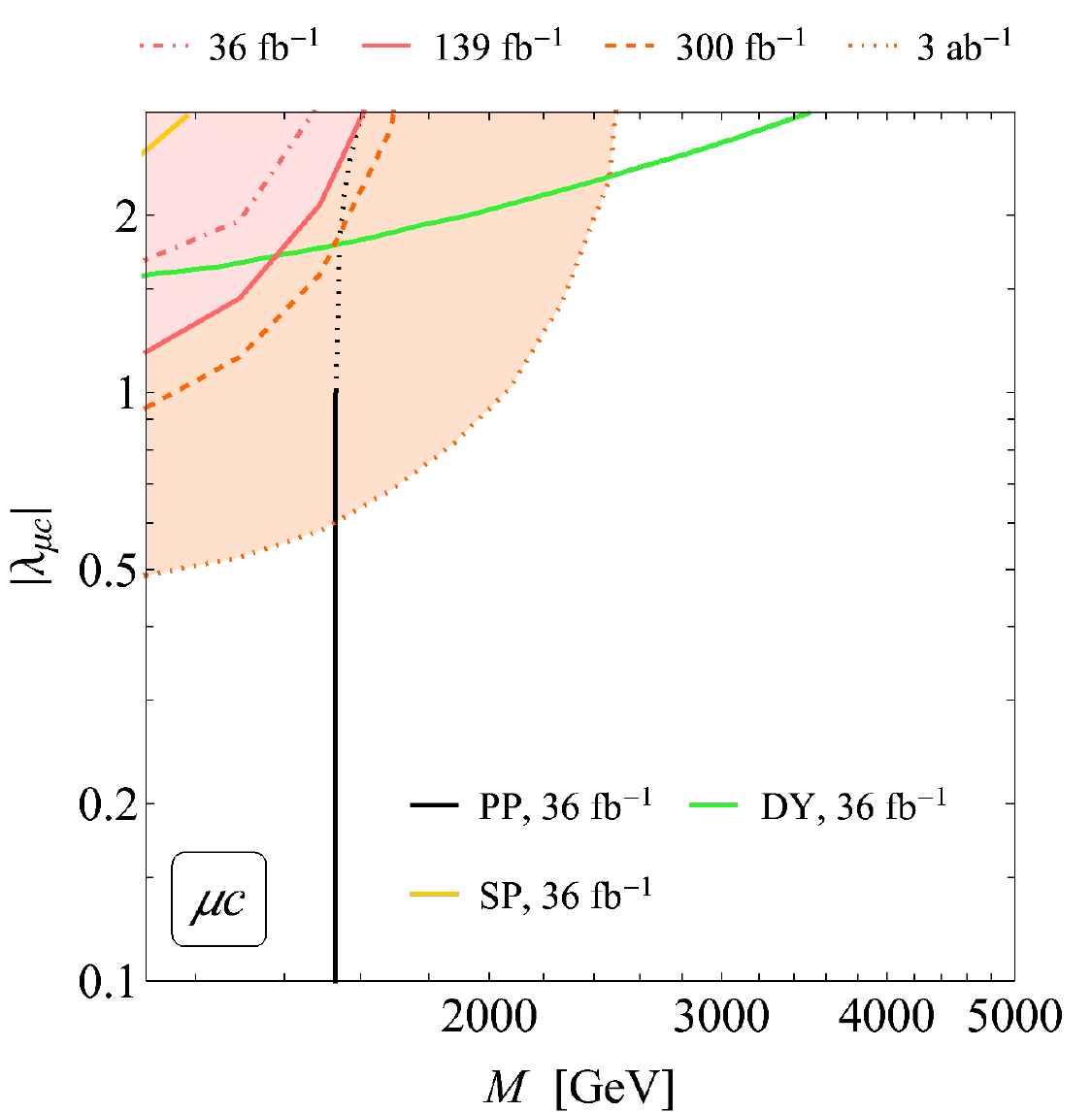} 
\vspace{2mm} 
\caption{\label{fig:limits2}  95\%~CL limits on the parameter space of minimal ${\rm LQ}_{\mu q}$ bosons with $q = u, d, s, c$.  Apart from the  PP exclusions that have been obtained  in~\cite{Schmaltz:2018nls} by a recast of~\cite{Sirunyan:2018ryt}, the colours, styles and meanings of the shown bounds resemble those of Fig.~\ref{fig:limits1}. Consult the main text for further explanations. }
\end{center}
\end{figure*}

Figs.~\ref{fig:limits1} and \ref{fig:limits2} display  the most relevant 95\% confidence level~(CL) limits on the magnitude of the couplings $\lambda_{\ell q}$ as a function of the mass~$M$  for  minimal scalar LQs. All possible flavour combinations involving an electron  or a muon  and an up~($u$), a down~($d$), a strange~($s$) or a~charm~($c$) quark are considered. The bounds that derive from our novel search strategy for resonant single LQ production are displayed as red~(LHC~Run~II constraints) and orange (projections) shaded regions. The  dashed-dotted red, solid red and dashed orange lines assume an integrated luminosity of $36 \, {\rm fb}^{-1}$, $139 \, {\rm fb}^{-1}$ and $300 \, {\rm fb}^{-1}$ for $pp$ collisions at $\sqrt{s} = 13 \, {\rm TeV}$, respectively, while the dotted orange lines assume $3 \, {\rm ab}^{-1}$ of $\sqrt{s} = 14 \, {\rm TeV}$~data. The most stringent limits on the mass of minimal first-~\cite{Sirunyan:2018btu} and second-generation~\cite{Sirunyan:2018ryt}  scalar LQs, obtained from pair-production~(PP) searches, are indicated as black lines. These limits are based on $36  \, {\rm fb}^{-1}$ of LHC~Run~II data and correspond to  $M > 1435 \, {\rm GeV}$ and $M > 1530 \, {\rm GeV}$ for first- and second-generation LQs, respectively. Notice that PP via $t$-channel exchange of a lepton has not been considered in the CMS analyses. As shown in~\cite{Schmaltz:2018nls} this simplification has, however, a  minor impact  for $|\lambda_{\ell q}| \lesssim 1$. Following~\cite{Schmaltz:2018nls} we include the lepton exchange contribution and indicate the PP bounds by dotted black lines for $|\lambda_{\ell q}| >1$. The green lines correspond to the Drell-Yan~(DY) bounds derived in~\cite{Schmaltz:2018nls}  from the CMS results~\cite{Sirunyan:2018exx}, while the yellow lines  depict the single production~(SP) projections~\cite{Schmaltz:2018nls}  of the CMS LHC~Run~I search~\cite{Khachatryan:2015qda}. Both sets of bounds assume  $36  \, {\rm fb}^{-1}$  of $pp$ collisions at $\sqrt{s} = 13 \, {\rm TeV}$. The couplings~$\lambda_{eq}$ with $q = u, d$ are also subject to the constraints arising from atomic parity violation, and from parity-violating electron scattering experiments that measure the  weak charge~($Q_W$) of protons and nuclei.  The relevant  95\%~CL bound  reads $|\lambda_{eq}| < 0.17 \hspace{0.5mm} M/{\rm TeV}$~\cite{Schmaltz:2018nls} and is shown in the upper two panels of Fig.~\ref{fig:limits1} as a dashed blue line. As is evident from these panels, in the case of $\lambda_{eu}$ ($\lambda_{ed}$) the hypothetical $139 \, {\rm fb}^{-1}$ bounds obtained from $s$-channel single LQ production are more stringent than the constraints from  $Q_W$ measurements for  $M \lesssim 3.2 \, {\rm TeV}$~($M \lesssim 2.1 \, {\rm TeV}$).  At the end of the HL-LHC the corresponding limits can be expected to surpass the bounds from  $Q_W$ measurements for   minimal scalar LQ masses up to around~$5.7 \, {\rm TeV}$~($4.2 \, {\rm TeV}$). Strong bounds  on the couplings $\lambda_{\mu u}$ and $\lambda_{\mu d}$ are also obtained using our method, while the minimal scalar LQ interactions involving a $s$ or a $c$ quark are more difficult to constrain given the suppression of the relevant quark PDFs.

{\bf Conclusions and outlook.} In this letter, we have demonstrated that lepton-initiated processes, which so far have been completely neglected experimentally, can be valuable probes of BSM physics  at hadron colliders.  We have, in particular, shown that $s$-channel single LQ production provides very sensitive direct tests of certain LQ-$\ell$-$q$ couplings at the LHC. Our new proposal takes advantage of the fact that the lepton PDFs in the proton are sufficiently large to yield measurable rates for $\ell q \to {\rm LQ} \to \ell q $ scattering  in $pp$ collisions. By studying final states with a high-$p_T$ electron or muon and a light-flavour jet, we have shown that a simple $E_{T, {\rm miss}}$ requirement combined with a lepton and jet veto  are  sufficient to suppress all relevant SM backgrounds to a level that allows for a successful bump hunt for LQ masses in the TeV range, potentially leading to a discovery. For the case of minimal  scalar LQs, we have performed a dedicated analysis considering all possible flavour combinations of first-  and second-generation leptons and quarks. We have found that the limits that can be derived from  LHC~Run~II data represent  the most stringent direct constraints to date on the interactions involving an up or~a~down quark. In~fact, in the case of the LQ-$e$-$u$ (LQ-$e$-$d$) coupling the obtained exclusions are so strong that they  even surpass  the indirect limits from  $Q_W$ measurements  for LQ masses below  $3.2 \, {\rm TeV}$~($2.1 \, {\rm TeV}$).  At~the HL-LHC  the latter bounds can be expected to be pushed up to approximately~$5.7 \, {\rm TeV}$~($4.2 \, {\rm TeV}$). Given the suppression of the relevant quark PDFs, minimal scalar LQ interactions involving a strange or a charm quark are more difficult to constrain using our search strategy.   In view of the simplicity of the proposed LQ signature and its discovery reach, we urge the ATLAS and CMS collaborations to perform dedicated resonance searches in final states featuring a single electron or muon and a single light-flavour jet in future LHC runs. 

We finally note that after some modifications the general search strategy proposed by us can also be applied  to other flavour combinations and/or other LQ quantum numbers and spins. A particularly interesting application in this context seems  to be the case of third-generation vector LQs that may be constrained by looking for a resonant signal in the $\tau b$ channel. Given the more complicated nature  of this final state, performing a dedicated analysis is, however, beyond the scope of this letter and thus left for future work. 

{\bf Acknowledgements.} UH thanks Jonas~Lindert and Giacomo~Polesello for helpful discussions.  We~thank Johannes~Bl{\"u}mlein for making us aware of \cite{Ohnemus:1994xf}. The research of LB is supported in part by the Swiss National Foundation under contracts 200020\_188464 and IZSAZ2\_173357. PN acknowledges support from Fondazione Cariplo and Regione Lombardia, grant 2017-2070, and from INFN.


%

\end{document}